# THE ANNEALING INDUCED EXTRAORDINARY PROPERTIES OF SI BASED ZNO FILM GROWN BY RF SPUTTERING

*JING LI[1,2], LEILEI DENG[2] AND SUNTAO WU[1]\**

[1.] Pen-Tung Sah MEMS Research Center, Xiamen University, Xiamen 361005, China
[2.] Department of Physics, Xiamen University, Xiamen 361005, China
\*E-mail: wst@xmu.edu.cn

## ABSTRACT

ZnO thin films were prepared by RF magnetron sputtering method on $SiO_2/Si(111)$ substrate at different ratio of $Ar/O_2$, which were annealed at different annealing temperatures to improve the films' quality. The effects of the ratio of $Ar/O_2$ and annealing treatment on the ZnO films' properties were discussed. The crystal structures and surface morphologies of ZnO thin films were characterized by X-ray diffraction (XRD) and scanning electron microscopy (SEM). The XRD results show that the as-deposited films have dominant c-axis orientation and the quality of ZnO films is better when grown with the gas ratio of $Ar/O_2$ is 1:1. After the annealing treatment, the quality of the ZnO thin films was improved. The annealing treatment induces the change of grain size and lattice expanding, which induced the strengthening of the band-edge emission at around 371 nm.

## 1. INTRODUCTION

There has been much work recently demonstrating wide-band-gap semiconductor materials, such as GaN, ZnO, SiC, and AlN due to their extensive application in optoelectronic devices. ZnO and its alloy are very promising materials in ultra violet (UV) and blue light-emitting devices. The band gap is 3.3 eV for ZnO and tuneable to 2.8 and 4 eV alloyed with Cd and Mg, respectively [1, 2]. A larger exciton binding energy of 60 meV than that of GaN (~25 meV) makes ZnO more useful for efficient laser UV applications. To date, many technologies have been applied to growth ZnO thin film, such as vapor-liquid-solid [3], pulse laser deposition [4], molecule beam evaporation, thermal evaporation, metal organic chemical vapor deposition [5], sputtering [6] and so on. It was well know that the film's quality and properties are strongly influenced by growth and treatment conditions. In this letter, we report the preparation of high quality ZnO nano polycrystalline films by radio-frequency (RF) magnetron sputtering technology followed by thermal treatment. The films' surface morphologies, structure and optical properties were characterized by using scanning electron microscopy (SEM), X-ray diffraction (XRD), and CL spectroscopy.

## 2. EXPERIMENTAL

The RF magnetron sputtering system was used to deposit the ZnO film. The bare silicon wafers with 300nm thickness buffered $SiO_2$ layers were used as substrates. Zn (99.99%) was used as the target, and the distance between target and substrate was 50 mm. The substrate was rotated with the speed of 20 r/min. The reactive chamber was first pumped down to about $1.33 \times 10^{-3}$ Pa. After that, a mixture of argon and oxygen was introduced, during the processing; the gas pressure is maintained at 1 Pa. The ratio of $Ar : O_2$ was changed. The RF power was kept at 200W, while the substrate temperature was kept at 200°C. To study the effects of annealing on the ZnO films' properties, the samples were annealed at temperature of 600°C, 800°C and 1000°C. The annealing time was kept the same as 30 min, and then cooled spontaneously.

## 3. RESULTS AND DISCUSSION

Fig. 1 shows X-ray diffraction patterns of ZnO films (as deposited) at the different ratio of $Ar/O_2$ in processing. Table 1 shows the calcualtion results of various parameters in all samples by Sherrer's formula and Bragg's equation. In the table, d is the calculating value of the samples, and D indicates the diameter of the grains and θ is the Bragg angle. From figure 1 we can find that all samples exhibit a strong 2θ peaks at about 34º, corresponding to the (002) peak of ZnO. Since the relative intensity of the ZnO(002) diffraction peak is significantly strong compared to the ZnO(103) peak, it can be assured that the c-axis oriented ZnO film is prepared. In the figure 1 and the table 1, it can be found that the ZnO(002) diffraction peak moves towards to lower angle with the partial of $O_2$ rising up. Meanwhile, the width at half maximum (FWHM) of ZnO(002) diffraction peak increases, and the relative intensity decrease. This result demonstrates that the quality of the ZnO films turns poor with the partial of $O_2$ increasing. From Sherrer's formula *D=Kλ/βcosθ* and Bragg's equation *2dsinθ=λ*, we can find that the space between two crystal surface becomes larger with the increasing of the partial of $O_2$, while the grain size decreases. This fact may be understood from that the oxygen has the higher electron capture ability and a lower sputtering yield than argon, which will decrease the density of plasmas [7]. It can be concluded that the quality of ZnO films is better when the ratio of Ar/$O_2$ is 1:1.

Table.1 Test results of various parameters in all samples

| Samples | Ar/$O_2$ | 2θ/(°) | d/nm | D/nm | FWHM/(°) |
|---|---|---|---|---|---|
| #1 | 1:1 | 34.10 | 0.262933 | 26.0 | 0.32 |
| #2 | 1:2 | 34.06 | 0.263233 | 21.3 | 0.35 |
| #3 | 1:3 | 34.05 | 0.263308 | 20.8 | 0.40 |
| #4 | 1:4 | 33.97 | 0.263910 | 15.7 | 0.53 |
| #5 | 1:5 | 34.00 | 0.263684 | 20.3 | 0.41 |

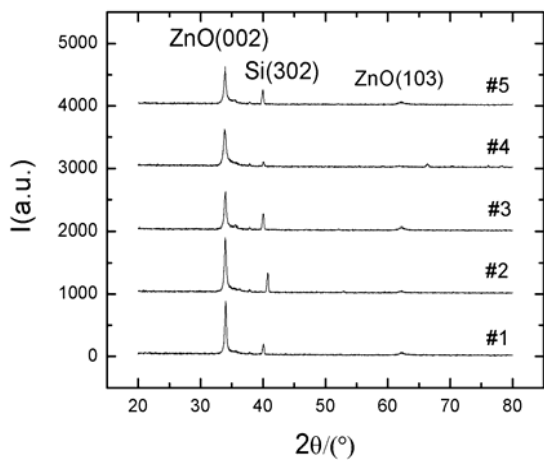

Fig.1 XRD spectra of ZnO films at the different ratio of Ar/$O_2$ in processing

As-deposited and annealed ZnO films were then imaged by SEM. The surface morphology of as-deposited ZnO film is characterized by a number of compact bright contrast spots in nano scale, as shown in Fig. 2(a). The average grain size is about 30 nm. No apparent changes of the surface morphologies are visible after the films are annealed in temperatures lower than 600°C, as illustrated in Fig. 2(b). Detail comparison with Fig. 2(a) show that grains become a little larger and more uniform as the film was annealed. When the annealing temperatures are higher than 600°C, the grains become distinct larger by connecting to neighbors and appear as a number of stacking flakes in sizes up to more than 100 nm, as shown in Fig. 2(c) and (d).

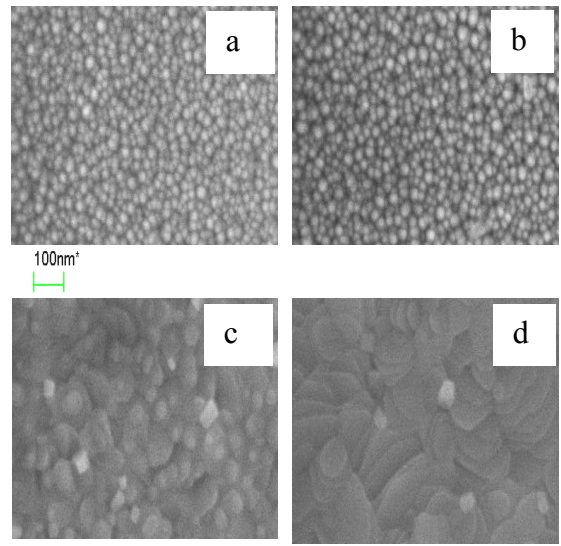

Fig.2 SEM images of the ZnO films at the different annealing temperature: (a)as-deposited, (b)600°C, (c)800°C and (d)1000°C.

To characterise the crystal structures of the ZnO films, XRD measurement was carried out. The XRD patterns of the films are shown in Fig. 3. All the diffraction peaks can be indexed to the wurtzite structure of ZnO. The dominant diffraction peak is around 34° which is assigned to the ZnO (002) plane, while a peak around 72° attributed to the (004) plane. Besides these, a very minor diffraction peak around 31° appears in samples annealed in 800 and 1000°C which is related to (100) plane.

Although all the ZnO films are crystalline and well orientated in *c*-axis, slight shift is visible for the (002) diffraction peak. It shifts from 34.21° (as-deposited) to 34.55°, 34.58°, and 34.63° for the samples annealed at 600°C, 800°C, and 1000 °C, respectively. Scherrer formula and Braag equation were applied to estimate the grain sizes and the lattice constant *c* and *a* for each sample. The SEM images and XRD results both show that the ZnO film by 1000°C annealing treatment ( so called ZnO-1000) has the large grain size with 54 nm and are most close to bulk ZnO in structure among the four samples. Comparing to the sample ZnO-1000, the grain size becomes smaller and the lattice constant *c* and *a* are increasing with decreasing the annealing temperature and without annealing. These results are also consistent with the TEM images which will be illustrated in latter paper. This may caused by the nano effect that smaller grain size will result in big surface influence. Moreover the lower diffusion and mobility of atoms at lower annealing temperature makes the crystal is not very ordered in *a* axis, although it has very strong *c*-axis orientation. It can be concluded that thermal treatment induces the change of the crystal structure in ZnO film.

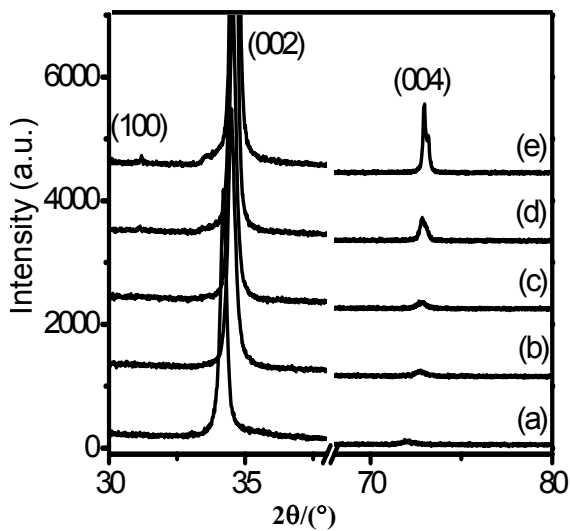

Figure 3. XRD pattern of ZnO film as deposited (a) and annealed at (b) 600°C, (c) 800°C, (d) 1000°C.

Fig. 4 shows the CL emission spectra of ZnO film without and with annealing treatment excited by 325 nm laser line at room temperature. No emission peak was detected in as deposited ZnO film and low temperature treated ZnO film at 600°C. While when the film annealed at higher temperature with 800°C and 1000°C, dominate emission peak at around 371 nm in both films, which was assigned to the band-edge emission of ZnO. Moreover, the band-edge emission becomes stronger when increasing the annealing temperature from 800°C to 1000°C, which is due to the improvement of the crystal structure of ZnO film at high temperature treatment. This result is consistent with the above XRD results. Besides the band-edge emission, a broad peak at around 550 nm ~ 620 nm is appeared in the films treated at 800°C and 1000°C, with the higher intensity at high temperature of 1000°C. This broad peak can be attributed to the defects in ZnO film such as interstitial zinc [8], or defects at grain boundaries [9].

## 4. CONCLUSION

In this work, nano ZnO films were grown by radio-frequency (RF) magnetron sputtering technology followed by thermal treatment. ZnO thin films in nano scale was prepared by sputtering method, which show extraordinary properties in XRD and CL spectra results due to nano-effect. On the films without annealing and with lower annealing temperature treatment at 600°C, the lattice constant *c* and *a* are larger than those films annealed at higher temperature of 800°C and 1000°C. Moreover, with the increasing of the annealing temperature, film's surface emission peak is strengthened.

## 5. ACKNOWLEDGEMENT

The authors gratefully acknowledge the financial support from Ministry of Science and Technology of China (No.2001CB610506), the Natural Science Foundation of China (No.90206039), Science and Technology Project of Fujian provice of China (No.2005H043), and Science and Technology Project of Xiamen University.

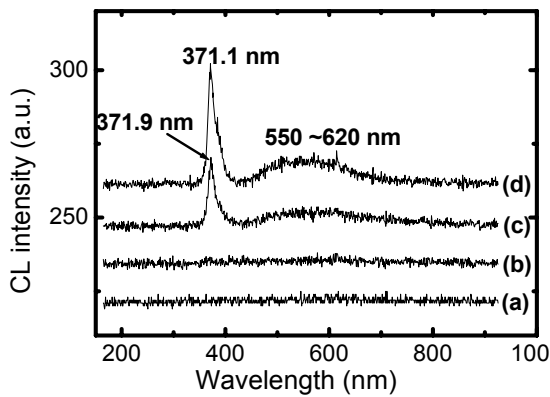

Figure 4. CL spectra of ZnO film excited by 325 nm laser as deposited (a) and annealed at (b) 600°C, (c) 800°C, (d) 1000°C.